\documentstyle[amssymb,12pt]{article}

\begin{document}

\title{{\bf Non-Invariant Ground States, Thermal Average, and Generalized Fermionic Statistics%
}}
\author{{\bf R. Mignani*}\thanks{%
e-mail: mignani@fis.uniroma3.it}{\bf and R. Scipioni}$^{\circ }$ \thanks{%
e-mail: scipioni@physics.ubc.ca{}}}
\maketitle

\begin{abstract}
We present an approach to generalized fermionic statistics which relates the existence
of a generalized statistical behaviour to non-invariant ground states.
Considering the thermal average of an operator generalization of the
Heisenberg algebra, we get an occupation number which depends on the degree
of mixing between symmetric and antisymmetric sectors of the ground state. A
natural prescription is given for the construction of a supersymmetric
statistics. We also show that the structure of the vacuum, and therefore the
statistical behaviour of the system, can be accounted for in terms of a
second-order phase transition.\newline
\newline
\end{abstract}

* Dipartimento di Fisica ''E. Amaldi'', Universit\`{a} degli Studi di Roma
''Roma Tre'', Via della Vasca Navale 84, 00146 Roma, Italy\newline
$^{\circ }$ Department of Physics and Astronomy, The University of British
Columbia,\newline
6224 Agricultural Road, Vancouver, B.C., Canada V6T 1Z1\\
\\
\bigskip Pacs: 05., 05.30.-d, 05.30. Pr \newpage In the last few years there
has been increasing interest in particles obeying statistics different from
Bose or Fermi. The early observation by Green [1] that different types of
statistics are allowed in the context of field theory  produced an enormous
lot of work on the so-called para-statistical fields [2]. The commutation
relations for these fields are trilinear in the creation and annihilation
operators and characterized by an integer $p$ which is the order of the
para-statistics. The case when $p$ is not an integer has been recently
studied [3] in order to provide theories in which the Pauli exclusion
principle and/or Bose statistics can be slightly violated.\newline
More recently another possibility has been considered, which corresponds to
the case where no assumption is made about the parameter $p$ [4-6].\newline
The idea is to consider ``quon'' creation and annihilation operators
satisfying the relations: 
\begin{equation}
a(f)a^{+}(g)-qa^{+}(g)a(f)=\left\langle f,g\right\rangle {\bf 1}
\end{equation}
where $f$ , $g$ are test functions, i.e. elements of the one-particle space
with inner product $\left\langle f,g\right\rangle $, and the quon operators $%
a(f)$, $\ a^{+}(f)$ depend linearly on $f$. As proved by Fivel, the
deformation\ (or mutation) parameter $q$ must be real in the interval $-1<q<1
$ [7].\newline
The quon system with a single degree of freedom, for which the test function
is one-dimensional, and only one relation $aa^{+}-qa^{+}a=1$ remains, is
called a $q$ oscillator. It was first introduced on a mathematical basis [8]
and later shown to naturally arise in the theory of quantum $SU(2)$ [9] by
Biedenharn and MacFarlane [10].\newline
Recently it has been observed that, by considering a 'quon' field model, the
locality which does not hold by virtue of Fredenhagen theorem [11] is
restored when a thermal average is considered, leaving room then for a
well-defined perturbation theory (at least at the two-point function level)
[6].\newline
On the other hand, the parameter $q$ which appears in [6] is not dynamical,
so it would be worth obtaining a model which interprets somehow $q$ as being
a dynamical parameter.\newline
In this letter, using an operator generalization of the Heisenberg-Weyl
algebra [12-14], we show how to relate, in the context of thermal field
theory and for the exclusion statistics 
the $q$ parameter to the symmetries of the ground state. We will
show that $q$ represents the degree of mixing between the 'bosonic' and
'fermionic' sectors of the vacuum state; moreover, this approach allows us
to give a natural 'supersymmetric' scheme for a generalized statistics.
\newline
The generalized fermionic Heisenberg algebra ({\it ''guon algebra''}) we start from is
[12-14]: 
\begin{eqnarray}
aa^{+}-\hat{g}a^{+}a=1 \\ \nonumber
a^2 = {a^{+}}^{2} = 0
\end{eqnarray}
where the {\it statistical operator }$\hat{g}$ is defined by 
\begin{equation}
\begin{array}{c}
\hat{g}^{2}=1 \\ 
\hat{g}=\hat{g}^{+} \\ 
\lbrack \hat{g},a]=0 \\ 
\lbrack \hat{g},a^{+}]=0
\end{array}
\end{equation}
The first of (2) has been also shown recently [15] to be equivalent to the normalized form of a Heisenberg algebra with reflection, for a special form and with a different meaning of $\hat{g}$.
Together with the algebra (2), (3) we require the following conditions to
hold:

\begin{eqnarray}
\lbrack N,a] &=&-a \\
\lbrack N,a^{+}] &=&a^{+}  \nonumber
\end{eqnarray}
where the number operator is defined as usual: $N=a^{+}a$, observe that we need to satisfy the condition $(1- \hat{g}) \, a^{2}$.

The multiparticle states are obtained by the relation: 
\begin{equation}
\left| n\right\rangle =\frac{(a^{+})^{n}}{\sqrt{n_{g}!}}\left| 0\right\rangle 
\end{equation}
Wher use has been made of all the relations apart from the 2nd in (2). The factorial $n_{g}!$ is defined as: $n_{g}!=n(n-1).....(1+q)$ where $q$ is the vacumm expectation value of $g$ and will calculated in the following (see comments after (16)). However if use is made of the second in (2) we get that only $n =0,1$ are allowed.\\
Note the interesting fact that the factorial is affected only trough the term $1+q$ which replaces $2$ in the usual expression
By its very definition, the operator $\hat{g}$ has eigenvalues $\pm 1$
.Then, a general vacuum state $\left| 0\right\rangle $ can be decomposed as: 
\begin{eqnarray}
\left| 0\right\rangle  &\equiv &\left| 0(\theta )\right\rangle =\cos
\,\theta \left| 0_{+}\right\rangle +\sin \,\theta \left| 0_{-}\right\rangle 
\\
\hat{g}\left| 0_{\pm }\right\rangle  &=&\pm \left| 0_{\pm }\right\rangle  
\nonumber
\end{eqnarray}
where $\left| 0_{\pm }\right\rangle $ denotes, respectively, the bosonic and
the fermionic vacuum, and\ the notation $0(\theta )$ is aimed at explicitly
stressing that {\it different }vacuum states are obtained for {\it different 
}values of the mixing angle $\theta $.

In the thermal field theory approach, we have to deal with statistical
averages which typically involve quantities of the kind: 
\begin{equation}
{\cal N}\,Tr(e^{-\beta H}TPol(X(t)))
\end{equation}
where ${\cal N}$ is a normalization constant, $\beta $ the inverse
temperature, $Pol(X(t))$ a polynomial in the position operator and $T$ a
prescription of time ordering.\newline

The thermal average of a quantity $Q$ over the fermionic sector is defined by 
\begin{equation}
\left\langle Q\right\rangle _{\beta }\equiv \frac{1}{Z}\sum_{0,1}\left\langle
n\right| e^{-\beta H}Q\left| n\right\rangle 
\end{equation}
where the Hamiltonian reads: 
\begin{equation}
H=\omega N
\end{equation}

By taking the thermal average of eq.(2) 
\begin{equation}
\left\langle aa^{+}\right\rangle _{\beta }-\left\langle \hat{g}%
a^{+}a\right\rangle _{\beta }=1
\end{equation}
it follows, from the commutation rules (4): 
\begin{equation}
e^{-\beta H}a^{+}=a^{+}e^{-\beta H}e^{-\beta \omega }
\end{equation}
so that 
\begin{equation}
\left\langle aa^{+}\right\rangle _{\beta }=e^{\beta \omega }\left\langle
a^{+}a\right\rangle _{\beta }
\end{equation}
and 
\begin{equation}
\lbrack a^{+}a,e^{-\beta H}]=0
\end{equation}
Then, it is easy to verify that: 
\begin{equation}
\left\langle \hat{g}a^{+}a\right\rangle _{\beta }=\left\langle
a^{+}a\right\rangle _{\beta }\cos \,2\theta 
\end{equation}
whence eq.(10) becomes 
\begin{equation}
(e^{\beta \omega }-\cos \,2\theta )\left\langle a^{+}a\right\rangle _{\beta
}=1
\end{equation}
or: 
\begin{equation}
\left\langle a^{+}a\right\rangle _{\beta }=\frac{1}{e^{\beta \omega }-\cos
\,2\theta }
\end{equation}

Two points are worth stressing. First, the thermal average of the number
operator depends on the angle $\theta $, which, by virtue of eq. (6),
characterizes the vacuum state. In general, a different value of $\theta $
corresponds to a different ground state and therefore to a different
statistics. Then, in this approach, {\it the generalised fermionic statistical behaviour depends
on the ground state.}

Moreover, let us notice that the above expression coincides with what we
would get considering the thermal average of the algebra (2) with $\hat{g}$
replaced by $q=\cos \,2\theta $. Thus, {\it for a fixed value of} $q$, the
two approaches are equivalent, but of course in our case $q(\theta )=\cos
2\theta $ is dependent on the ground state.\newline

Let us now show that our formalism permits to get in a natural way a
supersymmetric statistics. Under the supersymmetric transformation 
\begin{equation}
\left| 0_{+}\right\rangle \leftrightarrow \left| 0_{-}\right\rangle 
\end{equation}
eq.(16) goes into 
\begin{equation}
\left\langle a^{+}a\right\rangle _{\beta }=\frac{1}{e^{\beta \omega }+\cos
\,2\theta }
\end{equation}
To obtain a supersymmetric statistics requires to symmetrize (16) with
respect to the transformation (17). We get therefore 
\begin{equation}
\left\langle a^{+}a\right\rangle _{\beta }=\frac{1}{2}[\frac{1}{e^{\beta
\omega }-\cos \,2\theta }+\frac{1}{e^{\beta \omega }+\cos \,2\theta }]=\frac{%
e^{\beta \omega }}{e^{2\beta \omega }-(\cos \,2\theta )^{2}}
\end{equation}
The former expression can be rewritten as: 
\begin{equation}
\left\langle a^{+}a\right\rangle _{\beta }=\frac{1}{e^{\beta \omega }-\cos
\,2\tilde{\theta}}
\end{equation}
i.e. the same form of (16), with the angle $\tilde{\theta}$ being defined
by: 
\begin{equation}
\tilde{\theta}=\frac{1}{2}\,\arccos \left( \frac{\cos \,2\theta }{e^{\frac{%
\beta \omega }{2}}}\right) ^{2}
\end{equation}
In other words, {\it \ the supersymmetric statistics (19) is recovered as a
special case of our formalism}, and corresponds to the following choice of
the vacuum state 
\begin{equation}
\left| 0(\tilde{\theta})\right\rangle =\cos \,\tilde{\theta}\left|
0_{+}\right\rangle +\sin \,\tilde{\theta}\left| 0_{-}\right\rangle 
\end{equation}
with {\it temperature-dependent} mixing coefficients.\newline
\qquad Observe the following two limiting cases:\newline

\bigskip 

a) \underline{Zero temperature}:\newline
\newline
$\tilde{\theta}\rightarrow (\frac{\pi}{4} + k \frac{\pi}{2})$ with $k$ any integer; \ in this case, the supersymmetric system satisfies the condition $cos \, 2 \tilde{\theta} = 0$ then we get the \emph{Maxwell-Boltzmann} statistics.\newline

\bigskip 

b) \underline{Infinite temperature}:\newline
\newline
$\tilde{\theta}\rightarrow \frac{1}{2}\arccos \left( \cos \,2\theta \right)
^{2}$\newline
\newline

Let us notice that while it is true that the statistics (19) is invariant
under the supersymmetric transformation (17), it still depends on the
parameter $\theta $. Thus, in general we get again {\em different}
supersymmetric statistics for different values of $\theta $, i.e. for
different vacuum states. For instance, for a nearly pure bosonic (fermionic)
vacuum state $\theta \cong 0$ ($\theta \backsimeq \pi /2$), $\tilde{\theta}=%
\frac{1}{2}\,\arccos \left( e^{-\beta \omega }\right) $, while for an
equi-mixed vacuum $\theta =\frac{\pi }{4}$ and $\cos (2\tilde{\theta})=0$,
so that we get as in the zero temperature limit
\begin{equation}
\left\langle a^{+}a\right\rangle _{\beta }=e^{-\beta \omega }
\end{equation}
the {\em Maxwell-Boltzmann} statistics. In this sense {\it the MB
statistics may be considered as the 'maximally' supersymmetric statistics.}

We may ask whether it is possible, in our framework,  to obtain a statistics
which is ''{\it vacuum invariant''}, i.e. invariant under a generic rotation 
$U(\theta )$ . Clearly the simplest way to achieve this is to sum the
thermal averages of the number operators over all possible values of $\theta 
$.\newline
In the continuum case, and for supersymmetric statistics, this amounts to take: 
\begin{equation}
\overline{\left\langle a^{+}a\right\rangle _{\beta }}=\frac{1}{\pi }{%
\int_{0}^{\pi }}d\theta f(\theta )\frac{e^{\beta \omega }}{e^{2\beta \omega
}-(\cos 2\theta )^{2}}
\end{equation}
\bigskip This expression is invariant under a generic shift $\theta
\rightarrow \theta +\Delta \theta $ provided that $f(\theta )$ is periodic, $%
f(\theta )=f(\theta +\pi )$. Therefore, the sum (24) is a prescription
yielding a thermal-averaged occupation number which by construction does not
contain any $\theta $ dependence. So the previous relation defines a very
general supersymmetric, vacuum-invariant statistics obtained from (2) in the
context of thermal field theory.

A basic question within the present approach is: What does determine the
nature of the generalized statistics? Since, as we showed above, the
mutation parameter $q$ depends on the vacuum state (6) (through the mixing
angle $\theta $), this amounts to ask what determines the mixing between the
bosonic and fermionic ground states. A possible solution might be provided
by a symmetry-breaking mechanism, or a second-order phase transition,
depending on the critical values taken by some physical variables
characterizing the system. Although we deserve to face this problem in more
detail elsewhere, we briefly present a scheme for this mechanism.

First of all, notice that the guon algebra (2) admits (contrarily to the
quon algebra) the following freedom in the definition of the statistical
operator  $\hat{g}$ :
\begin{equation}
U(x)\hat{g}U^{+}(x)=\hat{g}^{\prime }
\end{equation}
where $U(x)$ is a group of local unitary transformations, acting on the
2-dimensional space spanned by the eigenstates of $\hat{g}$, whose action on
the state $\left| 0(\theta )\right\rangle $ is defined by
\begin{equation}
U(x)\left| 0(\theta )\right\rangle =\left| 0^{\prime }(\theta )\right\rangle 
\end{equation}
In   vector notation, we can write
\begin{equation}
\left| 0(\theta )\right\rangle =\left( 
\begin{array}{c}
\cos \theta  \\ 
\sin \theta 
\end{array}
\right) 
\end{equation}
and
\begin{equation}
U(x)=\left( 
\begin{array}{cc}
\cos \delta (x) & -\sin \delta (x) \\ 
\sin \delta (x) & \cos \delta (x)
\end{array}
\right) 
\end{equation}
so that 
\begin{equation}
U(x)\left| 0(\theta )\right\rangle =\left( 
\begin{array}{c}
\cos (\theta +\delta ) \\ 
\sin (\theta +\delta )
\end{array}
\right) 
\end{equation}
This allows us to define locally the mixing angle, and therefore the
parameter $q$:
\begin{equation}
q(x)=\cos 2\theta (x)\ 
\end{equation}
In general we may conceive the existence of models in which an order parameter $\eta (x)$ is introduced:
\begin{equation}
\eta (x)=F(x,q(x),T)
\end{equation}
where $T$ indicates the temperature and $x$ indicates a set of variables on which $q$ and $F$ depend.\\
The order parameter $\eta (x)$ is a scalar field, entering the second-order
phase transition Lagrangian  
\begin{equation}
L=a\eta ^{4}+b\eta ^{2}+c
\end{equation}
with $b=A(T-T_{c})$.

If we assume $T>T_{c}$, the minimum of $L$ is obtained for $\eta =0$, which
implies 
\begin{equation}
F(x,q(x), T_{+}) = 0
\end{equation}
From which we can obtain the values of $q(x)$. The specific expression in general will depend on the model we choose.\\
For $T<T_{c}$, the Lagrangian has two minima, for the following values of $%
\eta :$%
\begin{equation}
\eta _{\pm }=\pm \sqrt{\frac{A(T_{c}-T)}{2a}}
\end{equation}
Then we have to satisfy the equation:
\begin{equation}
F(x,q(x), T_{-}) = \pm \sqrt{\frac{A(T_{c}-T)}{2a}}
\end{equation}
From which in general we get two values of $q(x)$.\\
In the limit $T \longrightarrow 0$ , $\eta \longrightarrow \pm \sqrt{AT_{c}/2a}$.
From the limit discussed previously, we need to satisfy:
\begin{equation}
\eta_{\pm} = F(x_{\pm}, 0, 0)
\end{equation}

In conclusion, we have presented an approach to generalized fermionic statistics 
in which the statistical behaviour of a system is determined by the vacuum
structure. In this framework, the parameter $q$ is interpreted as the
thermal average of an operator $\hat{g}$ entering a generalized algebra
between creation and annihilation operators. Moreover, our formalism allows
us to obtain a supersymmetric statistics in a natural way using (20), and a
vacuum invariant one using (24). As an interesting example, it was shown
that the Maxwell-Boltzmann statistics originates from a supersymmetric
system of equi-mixed bosons and fermions.  We outlined how a phase transition-like behaviour  permits to account, at least in some cases, of the vacuum structure, and therefore of the generalized statistical behaviour.
\bigskip
\bigskip
\bigskip
{\bf ACKNOWLEDGMENTS}\\
\newline
One of us (RS) wish to thank the NOOPOLIS foundation (Italy) for partial
financial support and G. Semenoff for an interesting discussion.\newline

\end{document}